# The eBDIMS path-sampling server: generation, classification and interactive visualization of protein ensembles and transition pathways in 2D-motion space

Laura Orellana[1,*], Johan Gustavsson[2], Cathrine Bergh[3], Ozge Yoluk[3], and Erik Lindahl[1,3]*

[1] Science for Life Laboratory, Department of Biochemistry and Biophysics, Stockholm University, 106 91 Stockholm, Sweden
[2] Department of Computational Science and Technology, KTH Royal Institute of Technology, Lindstedtsv. 5, 100 44 Stockholm, Sweden
[3] Swedish e-Science Research Center, Dept. Applied Physics, KTH Royal Institute of Technology, Box 1031, 171 21 Solna, Sweden

* To whom correspondence should be addressed.

[L.O.] Tel: +46 727 091 468; Email: laura.orellana@scilifelab.se

[E.L.] Tel: +46 734 618 050; Email: erik@kth.se

Present Address: [O.Y.], Department of Pharmaceutical Sciences, School of Pharmacy, University of Maryland, Baltimore, Maryland, 21201, United States

## ABSTRACT

The recent rise of cryo-EM and X-ray high-throughput techniques is providing a wealth of new structures trapped in different conformations. Understanding how proteins transition between different conformers, and how they relate to each other in terms of function is not straightforward, and highly depends on the choice of the right set of degrees of freedom. Here we present eBDIMS server, an online tool and software for automatic classification of structural ensembles and reconstruction of transition pathways using coarse-grained (CG) simulations. The server generates CG-pathways between two protein conformations along with a representation in a simplified 2D-motion landscape based on the Principal Components (PCs) from experimental structures. For a conformationally rich ensemble, the PCs provide powerful reaction coordinates for automatic structure classification, detection of on-pathway intermediates and validation of *in silico* pathways. When the number of available structures is low or sampling is limited, Normal Modes (NMs) provide alternative motion axes for trajectory analysis. The path-generation eBDIMS method is available at a user-friendly website: https://login.biophysics.kth.se/eBDIMS/ or as standalone software. The server incorporates a powerful interactive graphical interface for simultaneous visualization of transition pathways in 2D-motion space and 3D-molecular graphics, which greatly facilitates the exploration of the relationships between different conformations.

## INTRODUCTION

Protein dynamics is essential to understand biological function. In spite of enormous technical advances in structural biology and molecular simulations, a central problem is to properly sample the conformational landscape of proteins, i.e. to detect the multiple flexible states that explain the mechanisms at the atomic level and how they relate to function at the cell, tissue and organism scale. The transient nature of these intermediates and the multiple routes in multidimensional energy landscapes make their computational elucidation challenging, while experimental information is usually scarce. The recent rise of cryo-EM and X-ray high-throughput techniques is providing a wealth of new structures trapped in different conformations (1, 2), but yet, understanding how they relate to each other is complicated and requires information on the pathways that connect the stable end-states. Typically, validation of *in silico* generated transition pathways and their possible biological relevance is complex, and requires prior knowledge about the system to define ad hoc reaction coordinates specific for each case (e.g. distances or angles between domains, structural elements, etc). Public web servers that generate protein transitions are common (3–5) but the presently available ones only evaluate the generated transition intermediates in terms of stereochemical quality, and cannot asses the biological relevance of a pathway or extract new insights in the context of available experimental information.

We have recently shown that the Principal Components (PC) of a "structurally rich" ensemble of structures can provide intrinsic powerful reaction coordinates for transition pathway analysis (6). Basically, when the solved structures available in the Protein Data Bank (PDB) for a given protein (and its close homologs) sample at least three different interconnected, on-pathway conformations (i.e. the ensemble contains at least one well-defined path intermediate), the PCs provide optimal collective variables to describe interconversion paths in the conformational landscape and can provide totally new insight on biological mechanisms or relationships between available structures not accessible otherwise. Projection analysis can also help to monitor the sampling by Molecular Dynamics (MD) simulations in comparison with solved experimental structures and eBDIMS paths.

To facilitate such analysis, we here present a user-friendly online tool built around the eBDIMS algorithm (6) which uses a hybrid elastic-network/Brownian Dynamics model to generate CG-pathways between two functional protein end-states (defined by the user). A key feature of the eBDIMS web application is the clustering and analysis of protein structures by their projection onto the intrinsic normal mode (NM) and principal component (PC) motion-space. The tool classifies structures and pathways onto the motion space defined by the PCs derived from a structural ensemble of distinct conformations (provided by the user). Since accurate ensemble preparation may require manual curation and often there are not enough conformers solved for a protein, the eBDIMS web server also presents as default reaction coordinates the first two NMs of the starting structure computed by the ED-ENM force-field (7), which was calibrated against a MD library (8, 9) and experimental data from X-ray and NMR, and has been used to successfully assess CASP predictions (10). Low frequency NMs have been shown not only to describe efficiently conformational changes involving large rigid-body motions (5, 11, 12) but also to correlate with PCs from experimental ensembles or MD simulations (7, 13, 14). Therefore, they provide alternative axes useful to monitor how a given transition is proceeding in terms of molecular movements in case not enough representative conformers are sampled in the ensemble. In summary, each PC or NM describes a fundamental molecular movement, which often correlate (6) with the heuristic reaction coordinates typically described for a given protein, but avoid the ambiguity of user and system-tailored definitions.

We have thoroughly illustrated the functional insights (6) gained by this analysis approach for biomolecular systems of large size and complexity, comprising proteins from ion channels to nucleolytic enzymes. To the best of our knowledge, this is the first server combining path-sampling and PCA in an interactive and user-friendly interface, which can be of interest for a large community of experimental and computational biologists for the rationalization and integration of structural data.

## MATERIAL AND METHODS

eBDIMS transition pathways are calculated in both directions between two end-state protein structures (start and target, respectively), and subsequently projected onto an intrinsic 2D-motion space for visual inspection. The motion spaces are defined by NMs (default) or PCs (optional).

### The eBDIMS algorithm

The eBDIMS algorithm has been described elsewhere (6). Briefly, it is based on the ED-ENM (Essential Dynamics-refined Elastic Network Model) force field (7). It reduces the protein structure to a simplified bead-spring model where residues represented by positions of their C-alpha atoms are interconnected via harmonic springs to mimic covalent and electrostatic bonding. The ED-ENM potential energy (*U*) for a given atom (*i*) is given by

$$U_i = \frac{1}{2}\sum_{j=1}^{N} K_{ij}\left(r_{ij} - r_{ij}^{0}\right)^2 \qquad (1)$$

where *N* is the number of residues, $r_{ij}$ is the distance between C-alpha carbons *i* and *j*, and $K_{ij}$ is the force constant defined by the ED-ENM potential (7), with a distance cutoff for long term interactions set to 10Å. The elastic network is subject to an overdamped Langevin simulation (12, 15, 16), where random forces, $\xi$, play the role of thermal fluctuations and a friction coefficient, $\gamma$, is used to represent the implicit solvent:

$$m\ddot{\boldsymbol{r}}_i = \boldsymbol{F}_i - \gamma\dot{\boldsymbol{r}}_i + \boldsymbol{\xi}(t) \qquad (2)$$

Numerical integration of eq. 2 is done using the *Verlet* algorithm, and the transition progress is monitored at every certain number of unbiased steps (*k*) by computing a progress variable ($\Gamma$), which is defined as the difference in pairwise distances between the current structure and the target, *t*:

$$\boldsymbol{\Gamma} = \sum_{i,j=1}^{N}\left(d_{ij} - d_{ij}^{t}\right)^2 \qquad (3)$$

If the progress variable (eq. 3) decreases compared to the previous step, the current structure is accepted. Iteration proceeds until convergence is achieved in 99.9% or the time limit is exceeded (4h). Further details on eBDIMS, Langevin and ED-ENM algorithms are available elsewhere (6, 7, 12)

**Pathway projections onto the motion space**

The calculated transition pathways are projected onto an intrinsic 2D-motion space defined by the PCs of the ensemble structures (if provided) and the NMs of the start structure.

*Principal components and Normal Modes.* PCA is a statistical technique to extract dominant features from noisy data, widely applied to analyze MD simulations (17). For our purpose, the calculation of the PC-motion space requires a "structurally rich" ensemble (6) (i.e. coordinate sets describing at least three different conformations of the protein, including the start and target structures). Protein structures are aligned to that selected as start conformation in order to compute a 3N-dimensional covariance matrix, *C*, which describes the mean-square deviations in atomic coordinates from their mean position (diagonal elements) and the correlations between their pairwise fluctuations (off-diagonal elements). Diagonalization of this symmetric matrix *C* yields a set of eigenvectors and eigenvalues representing the motions that explain the variation in the atomic coordinates ordered by their variance; see further details elsewhere (6, 7, 13, 18). Similarly, diagonalization of the ED-ENM potential energy or "stiffness" matrix defined in (1) renders a set of eigenvectors and eigenvalues of increasing frequency that describe protein intrinsic motions (see (7, 12, 19)). In-house tools written in C++ and FORTRAN are used to perform PCA and NMA calculations.

The largest variance PCs and lowest frequency NMs typically capture the large-scale functional motions of a protein. Within this framework, each structure can be characterized by its projections onto the conformational space defined by two major motion components (NMs or PCs), $M_k$ *(k=1,2)*:

$$M_k = |\Delta r_{i-0}|.cos(M_k.r_{i-0}) \quad (4)$$

Where $\Delta r_{i\text{-}0} = (r - r_0) / \| r - r_0 \|$ is the 3N-dimensional unitary difference vector between the coordinates of the *i*-structure and the chosen reference, and $M_k$ is one of the two $k^{th}$ major axes.

The reliability of the NM-projections on each axis is estimated as the dot product between the end-states (0=start, t=target) transition vector and a given motion eigenvector $M_k$:

$$\alpha_k = \frac{\Delta r_{t-0} \cdot M_k}{\|\Delta r_{t-0}\| \|M_k\|} \quad (5)$$

Generalization of (6) for the *m*-important deformation modes yields to a similarity index ranging from 0 (no similarity in the directions of motion) to 1 (perfect similarity). For PCs, the two first PCs, which typically capture >70% of the variance for an ensemble (6, 7), are considered for projection.

## DESCRIPTION OF THE WEB APPLICATION

### Input data requirements

*Job name.* Each new job should be assigned a describing (alphanumeric) job name. If an email is provided the user receives a notification on job completion including a link to access the data. A graphical description of the eBDIMS server workflow is presented in *Fig.1*.

*End-state structures.* The eBDIMS web server requires two distinct end-state conformations of the same protein (referred to as start and target, respectively). End-state structures can be manually uploaded as coordinate files in strict PDB-format (with or without header), or retrieved by entering their PDB structure IDs (e.g. 4NPQ:A, to specify chain A of PDB 4NPQ). Importantly, input structures should not have any missing residues, and sequence should match perfectly between them. If end-state structures differ in number of chains, only the number of chains in the shorter PDB structure will be accounted for.

*Conformational Ensemble.* Input end-state files can be accompanied by a third input file that contains several coordinate sets perfectly matching in residue number with the start and target structures. For an experimental ensemble, it is recommended that the structures correspond to as many different conformations as possible, preferably distinct from the end-states. It is also possible to upload MD simulations for inspection (<1000 frames). The coordinate sets must be delimited by MODEL and ENDMDL labels (i.e multi-pdb format) and should include the start and target end-state structures. Importantly, each coordinate set must be complete (i.e. no missing residues) and have the same number of residues as all other coordinate sets.

### Output results

After successful job submission the user is redirected to a unique result page, which displays the job status (i.e. pending/active/finished) together with information on the curated input files. The server performs all of the following functions:

*Generation of forward and reverse transition pathways*

Active jobs can be monitored in real-time from the result page via the status bar at the top of the page and in charts that regularly update the current transition progress (%) and RMSD between current structure and the target structure (Å). A typical simulation run takes from a few minutes to hours, depending on protein size and the complexity of the conformational change.

*Pathway projection and visualization onto a 2D-motion space*

Results are always presented in interactive 2D-motion projection plots in combination with 3D time-resolved molecular representations. The projection plots show the two eBDIMS transitions (*forward*:

start → target; *reverse*: target → start) and ensemble structures using two representative axes derived from NMA of the reference structure and PCA of an ensemble, if that has been provided. Ideally, each axis represents a biologically relevant movement for the transition, computed from single-structure NMA or directly from ensemble PCA and that can be visualized in a 3D-animation. The projection plots are further coupled to the 3D time-resolved molecular representations, so that it is also possible to visualize at the same time the generated pathway and their projections onto a simplified 2D-motion space, as well as its relationship with the uploaded structures.

*Default Projection onto NM-motion-space:* The server computes NMs from the start reference structure, and selects as projection axes the two modes that best overlap with the requested transition. The server reports: i) single overlaps of these NMs with the transition vector, and ii) cumulative overlap of the ten lowest-frequency NMs with the transition, which estimates the NM-space reliability as a representation of the conformational change. The closer to one (1.0) is the overlap the better is the NMA description of the transition. A low value (<0.20) indicates a random similarity and thus, that the axis/motion is not meaningful, while high values (>0.50) are on the contrary, significant.

*Optional Projections onto PC-motion-space:* When an ensemble is provided, the server computes the PCA and selects the two first PCs (those with the highest variance) as projection axes, which typically also display the best overlap with conformational changes. Tools like Bio3D (20) or ProDy (21) can help the user to prepare an ensemble from experimental structures. The quality of axes can be evaluated from their ability to cluster and classify structures according to their functional state. As a general guideline, obtaining consistently similar projections upon changes in ensemble composition and efficient clustering of related structural groups (for example, structures solved in complex with similar molecules) indicates reliable motion decomposition. In the case of MD, the relevance of the PCs will depend on the extent of the sampling achieved in the simulation, which can be appreciated in relation to the projections of the end-states as well as the eBDIMS paths.

## EXAMPLES OF USE CASES

Results generated by the eBDIMS server are provided for several highly representative proteins: ribose-binding protein (RBP), 5’-NTase, RNAseIII, SERCA, GLIC and Calmodulin. These proteins are excellent benchmark examples since they all have resolved structures in at least three different configurations. Moreover, the biological interpretation of the generated results has been thoroughly discussed previously (6). Further examples for other proteins with (*Table S1*) or without ensembles are available on the website. Importantly, listed examples provide access to the curated input data files (i.e., start, target, and ensemble PDB structures), which can be used to reproduce the results or serve as templates to prepare new job submissions for other proteins. Below we describe three representative situations.

**Structurally rich ensemble with harmonic NMA-sampled transition**

RBP is a bacterial protein that binds ribose with a 6Å hinge motion of two similar domains, sampled by a crystallographic ensemble (eleven structures) that covers the entire closing process, including intermediates (i.e. a structurally rich ensemble, as defined in (6)). For RBP, the unbound structure (PDB: 1BA2) can be selected as the reference for NMA calculations and PCA alignment. Typically, unbound structures are more open and have less interactions, generating NMs that capture better conformational changes (6, 22). By default, the lowest-frequency RBP modes (*Table S1*) are computed for projection: the dominant mode corresponds to a hinge opening-closing motion of the two globular domains ($O_1$=0.83) while the second tracks an orthogonal shear motion ($O_2$=0.40), which can be visualized in 3D-movies. This harmonic transition is perfectly described by the normal modes ($O_{10}$=0.97). Therefore, in this case projections onto the NM-space are capable of clustering the structures onto the three relevant conformations (open, intermediate and closed) (See *Fig.2 top right*). When the X-ray ensemble is provided, the server computes the PCs for projection analysis (See *Fig.2 top left*), which for RBP are very similar to NMs, with PC1 tracking bending and PC2, shear motion, also available in 3D-movies. In general, whenever PC projections are capable of clearly discriminating intermediate states, as shown here, they provide a better assessment of eBDIMS pathways and allow a better validation in terms of experimentally known intermediates.

Binding of ribose to RBP drives the transition towards the closed form. In about 4min, the forward (open to closed) and reverse (closed to open) transitions converge from 6 to 0.3 Å (*Table S1*), as can be monitored in real-time rMSD progress plots. Upon completion, the pathways can be explored interactively as 3D-movies coupled to projections onto the 2D-motion spaces (*Fig.2*), which allow immediate evaluation of its feasibility in terms of approaching intermediates and sampling smoothness. For this harmonic transition, both PCs and NMs cluster neatly the three conformations sampled along the eBDIMS forward pathway, and thus yield a similar representation of the transitions. The 2D-motion plots allow immediate identification of the structural intermediates, which appear as points spontaneoulsy visited along eBDIMS pathways.

**Structurally rich ensemble with complex conformational change**

RNaseIII is a homodimer that undergoes a complex concerted motion, sampled by a crystallographic ensemble (eleven structures) including several intermediates. The transitions spanning 17Å converge to 0.9 Å in less than 15 min (*Table S1*). Opposed to the former case, the lowest-frequency modes of RNAseIII (*Table S1*) do not track efficiently this complex transition ($O_{10}$=0.60): the major mode (NM1) captures the separation of RNA-binding domains ($O_1$=0.29) while the second (NM9) tracks a concerted rotation ($O_2$=0.39). Therefore, in this case the projections onto the NM-space, although clustering different conformations (open, intermediates and closed), are not capable of separating them clearly (See *Fig.2 bottom*), while in PC-space the relationships of eBDIMS pathways with structural intermediates become evident, with both the forward and reverse pathways approaching spontaneously two structural clusters. This is an illustrative example of a non-harmonic transition, in which NM-space is of limited use, while PC-space allows for pathway validation and intermediate classification, as well as for decomposition of a very complex motion sequence. However, the NM-space projections still allow to appreciate the asymmetry of the forward and reverse pathways, which is relevant for the catalytic mechanism (6).

## TECHNICAL DETAILS

The web server implementation relies on backend modules written in C, Python, bash, and PHP. The eBDIMS code has been parallelized with OpenMP to be capable of shared memory multi-threading. They run on hardware that consists of one Quad-Core AMD Opteron(tm) 64-bit Processors 2374 HE running at 2.20 GHz (8 cores) and equipped with 16 GB of RAM. Job submissions are managed and scheduled linearly. Presentation of results in interactive 2D-plots and as molecular 3D-representations is generated in the web browser using Javascript libraries d3.js and NGL viewer (23), respectively. Full documentation, tutorial and pre-computed examples, is provided.

## CONCLUSIONS

The first publicly available server for simultaneous pathway generation and analysis within the context of structural ensembles is presented here. The eBDIMS web server is an online tool that generates transition pathways between protein end-states, and allows for validation and rational analysis within the conformational landscape defined from multiple structures. The eBDIMS algorithm has been thoroughly validated with high quality experimental data, and we have shown that it can predict transition pathways and their intermediates with the same accuracy as fully atomistic force field, but with a fraction of computational cost, being superior to typically used morphing and path-sampling methods. The user-friendly interface can be useful for users without expert knowledge on structural bioinformatics, allowing for an intuitive visualization of the pathways generated and their relationships with multiple structures. Overall, we believe the eBDIMS server will be an important tool to classify new experimental structures and generate realistic pathways, which can serve as a starting point for structure reconstruction and more sophisticated analysis.

## AVAILABILITY

The eBDIMS source code is available under a GNU license in the GitHub repository:
https://github.com/langelion/eBDIMS

## ACCESSION NUMBERS

PDB IDs: 1BA2, 2DRI, 1YYO, 1YYW

## SUPPLEMENTARY DATA

A Supplementary Data file containing *Table S1* is available at NAR online.

## REFERENCES

1. Levantino,M., Yorke,B.A., Monteiro,D.C., Cammarata,M. and Pearson,A.R. (2015) Using synchrotrons and XFELs for time-resolved X-ray crystallography and solution scattering experiments on biomolecules. *Curr. Opin. Struct. Biol.*, **35**, 41–8.
https://doi.org/10.1016/j.sbi.2015.07.017
http://www.ncbi.nlm.nih.gov/pubmed/26342489

2. Nogales,E. and Scheres,S.H.W. (2015) Cryo-EM: A Unique Tool for the Visualization of Macromolecular Complexity. *Mol. Cell*, **58**, 677–689.
https://doi.org/10.1016/j.molcel.2015.02.019
http://www.ncbi.nlm.nih.gov/pubmed/26000851

3. Das,A., Gur,M., Cheng,M.H., Jo,S., Bahar,I. and Roux,B. (2014) Exploring the Conformational Transitions of Biomolecular Systems Using a Simple Two-State Anisotropic Network Model. *PLoS Comput. Biol.*, **10**.
https://doi.org/10.1371/journal.pcbi.1003521
http://www.ncbi.nlm.nih.gov/pubmed/24699246

4. Krüger,D.M., Ahmed,A. and Gohlke,H. (2012) NMSim web server: Integrated approach for normal mode-based geometric simulations of biologically relevant conformational transitions in proteins. *Nucleic Acids Res.*, **40**, 310–316.
https://doi.org/10.1093/nar/gks478
http://www.ncbi.nlm.nih.gov/pubmed/22669906

5. López-Blanco,J.R., Aliaga,J.I., Quintana-Ortí,E.S. and Chacón,P. (2014) iMODS: internal coordinates normal mode analysis server. *Nucleic Acids Res.*, **42**, W271–W276.
https://doi.org/10.1093/nar/gku339
http://www.ncbi.nlm.nih.gov/pubmed/24771341

6. Orellana,L., Yoluk,O., Carrillo,O., Orozco,M. and Lindahl,E. (2016) Prediction and validation of protein intermediate states from structurally rich ensembles and coarse-grained simulations. *Nat. Commun.*, **7**, 12575.
https://doi.org/10.1038/ncomms12575
http://www.ncbi.nlm.nih.gov/pubmed/27578633

7. Orellana,L., Rueda,M., Ferrer-Costa,C., Lopez-Blanco,J.R., Chacón,P. and Orozco,M. (2010) Approaching elastic network models to molecular dynamics flexibility. *J. Chem. Theory Comput.*, **6**, 2910–2923.
https://doi.org/10.1021/ct100208e
http://www.ncbi.nlm.nih.gov/pubmed/26616090

8. Meyer,T., D'Abramo,M., Hospital,A., Rueda,M., Ferrer-Costa,C., Pérez,A., Carrillo,O., Camps,J., Fenollosa,C., Repchevsky,D., *et al.* (2010) MoDEL (Molecular Dynamics Extended Library): a database of atomistic molecular dynamics trajectories. *Structure*, **18**, 1399–1409.
https://doi.org/https://doi.org/10.1016/j.str.2010.07.013 show
http://www.ncbi.nlm.nih.gov/pubmed/21070939

9. Rueda,M., Ferrer-Costa,C., Meyer,T., Pérez,A., Camps,J., Hospital,A., Gelpí,J.L. and Orozco,M. (2007) A consensus view of protein dynamics. *Proc. Natl. Acad. Sci. U. S. A.*, **104**, 796–801.
https://doi.org/10.1073/pnas.0605534104
http://www.ncbi.nlm.nih.gov/pubmed/17215349

10. Perez,A., Yang,Z., Bahar,I., Dill,K.A. and MacCallum,J.L. (2012) FlexE: Using Elastic Network Models to Compare Models of Protein Structure. *J Chem Theory Comput*, **8**, 3985–3991.
https://doi.org/10.1021/ct300148f
http://www.ncbi.nlm.nih.gov/pubmed/25530735

11. Bastolla,U. (2014) Computing protein dynamics from protein structure with elastic network models. *Wiley Interdiscip. Rev. Comput. Mol. Sci.*, **4**, 488–503.
https://doi.org/10.1002/wcms.1186

12. Orozco,M., Orellana,L., Hospital,A., Naganathan,A.N., Emperador,A., Carrillo,O. and Gelpí,J.L. (2011) Coarse-grained Representation of Protein Flexibility. Foundations, Successes, and Shortcomings. *Adv. Protein Chem. Struct. Biol.*, **85**, 183–215.
https://doi.org/https://doi.org/10.1016/B978-0-12-386485-7.00005-3
http://www.ncbi.nlm.nih.gov/pubmed/21920324

13. Yang,L.-W., Eyal,E., Bahar,I. and Kitao,A. (2009) Principal component analysis of native ensembles of biomolecular structures (PCA_NEST): insights into functional dynamics. *Bioinformatics*, **25**, 606–14.

https://doi.org/10.1093/bioinformatics/btp023
http://www.ncbi.nlm.nih.gov/pubmed/19147661

14. Yang,L., Song,G., Carriquiry,A. and Jernigan,R.L. (2008) Close correspondence between the motions from principal component analysis of multiple HIV-1 protease structures and elastic network modes. *Structure*, **16**, 321–30.
https://doi.org/10.1016/j.str.2007.12.011
http://www.ncbi.nlm.nih.gov/pubmed/18275822

15. Carrillo,O., Laughton,C.A. and Orozco,M. (2012) Fast Atomistic Molecular Dynamics Simulations from Essential Dynamics Samplings. *J. Chem. Theory Comput.*, **8**, 792–799.
https://doi.org/10.1021/ct2007296
http://www.ncbi.nlm.nih.gov/pubmed/26593340

16. Emperador,A., Carrillo,O., Rueda,M. and Orozco,M. (2008) Exploring the suitability of coarse-grained techniques for the representation of protein dynamics. *Biophys. J.*, **95**, 2127–2138.
https://doi.org/https://doi.org/10.1529/biophysj.107.119115 show
http://www.ncbi.nlm.nih.gov/pubmed/18487297

17. David,C.C. and Jacobs,D.J. (2014) Principal component analysis: a method for determining the essential dynamics of proteins. *Methods Mol. Biol.*, **1084**, 193–226.
https://doi.org/doi: 10.1007/978-1-62703-658-0_11

18. Katebi,A.R., Sankar,K., Jia,K. and Jernigan,R.L. (2015) The use of experimental structures to model protein dynamics. *Methods Mol. Biol.*, **1215**, 213–36.
https://doi.org/10.1007/978-1-4939-1465-4_10
http://www.ncbi.nlm.nih.gov/pubmed/25330965

19. Hayward,S. and Groot,B.L. De Normal Modes and Essential Dynamics. *Proteins*, **443**, 443.
https://doi.org/doi: 10.1007/978-1-59745-177-2_5
http://www.ncbi.nlm.nih.gov/pubmed/18446283

20. Grant,B.J., Rodrigues, a. P.C., ElSawy,K.M., McCammon,J. a. and Caves,L.S.D. (2006) Bio3d: an R package for the comparative analysis of protein structures. *Bioinformatics*, **22**, 2695–2696.
https://doi.org/10.1093/bioinformatics/btl461

http://www.ncbi.nlm.nih.gov/pubmed/16940322

21. Bakan,A., Meireles,L.M. and Bahar,I. (2011) ProDy: protein dynamics inferred from theory and experiments. *Bioinformatics*, **27**, 1575–7.
https://doi.org/10.1093/bioinformatics/btr168
http://www.ncbi.nlm.nih.gov/pubmed/21471012

22. Tama,F. and Sanejouand,Y.H. (2001) Conformational change of proteins arising from normal mode calculations. *Protein Eng.*, **14**, 1–6.
https://doi.org/https://doi.org/10.1093/protein/14.1.1
http://www.ncbi.nlm.nih.gov/pubmed/11287673

23. Rose,A.S. and Hildebrand,P.W. (2015) NGL Viewer: A web application for molecular visualization. *Nucleic Acids Res.*, 10.1093/nar/gkv402.
https://doi.org/10.1093/nar/gkv402
http://www.ncbi.nlm.nih.gov/pubmed/25925569

## FIGURES

# Figure 1

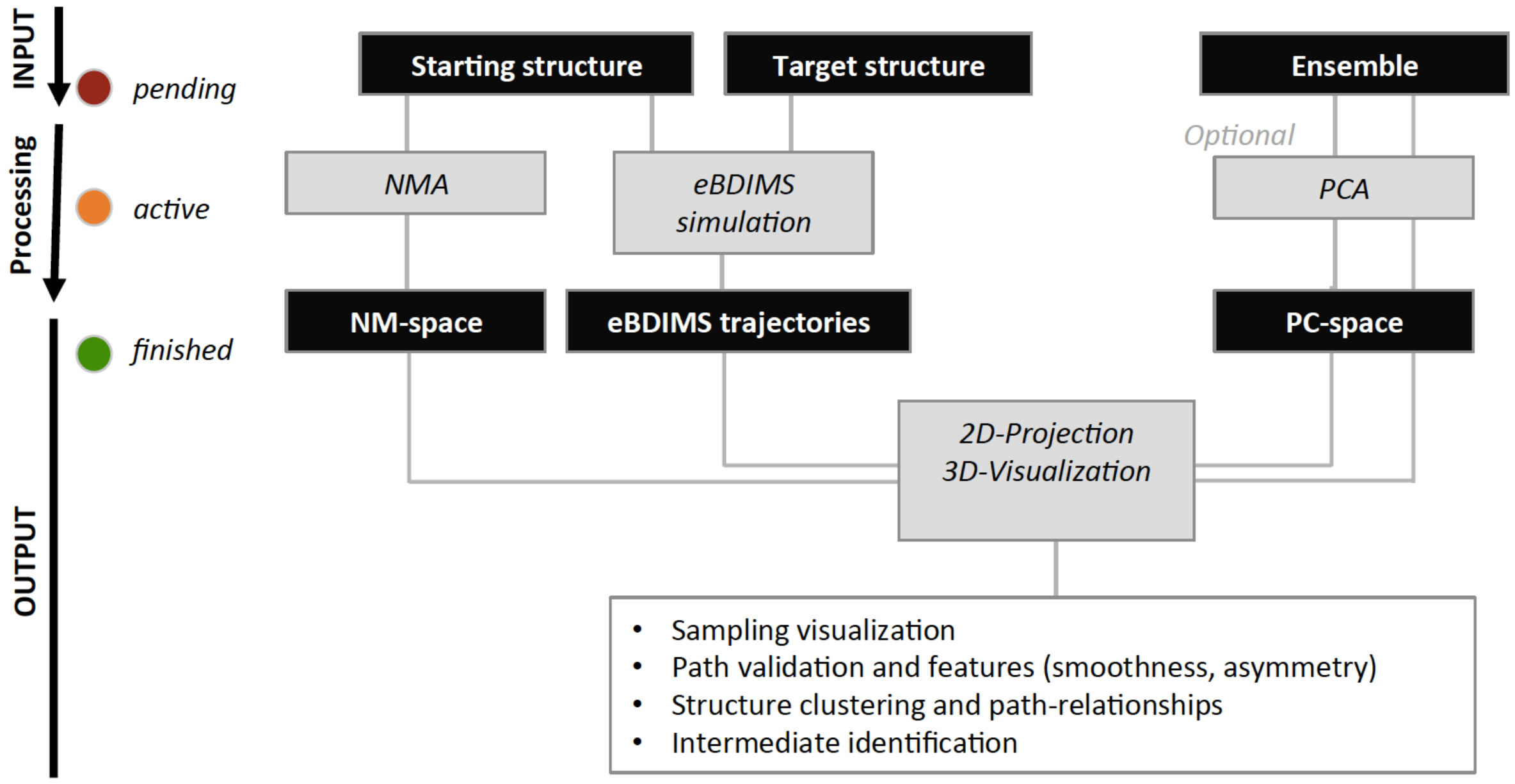

**Figure 1.** Flowchart of the EBDIMS webserver illustrating the input requirements (end-states and ensemble pdb files), the calculations performed and the output obtained (eBDIMS trajectories and projections onto 2D-motion space).

## Figure 2

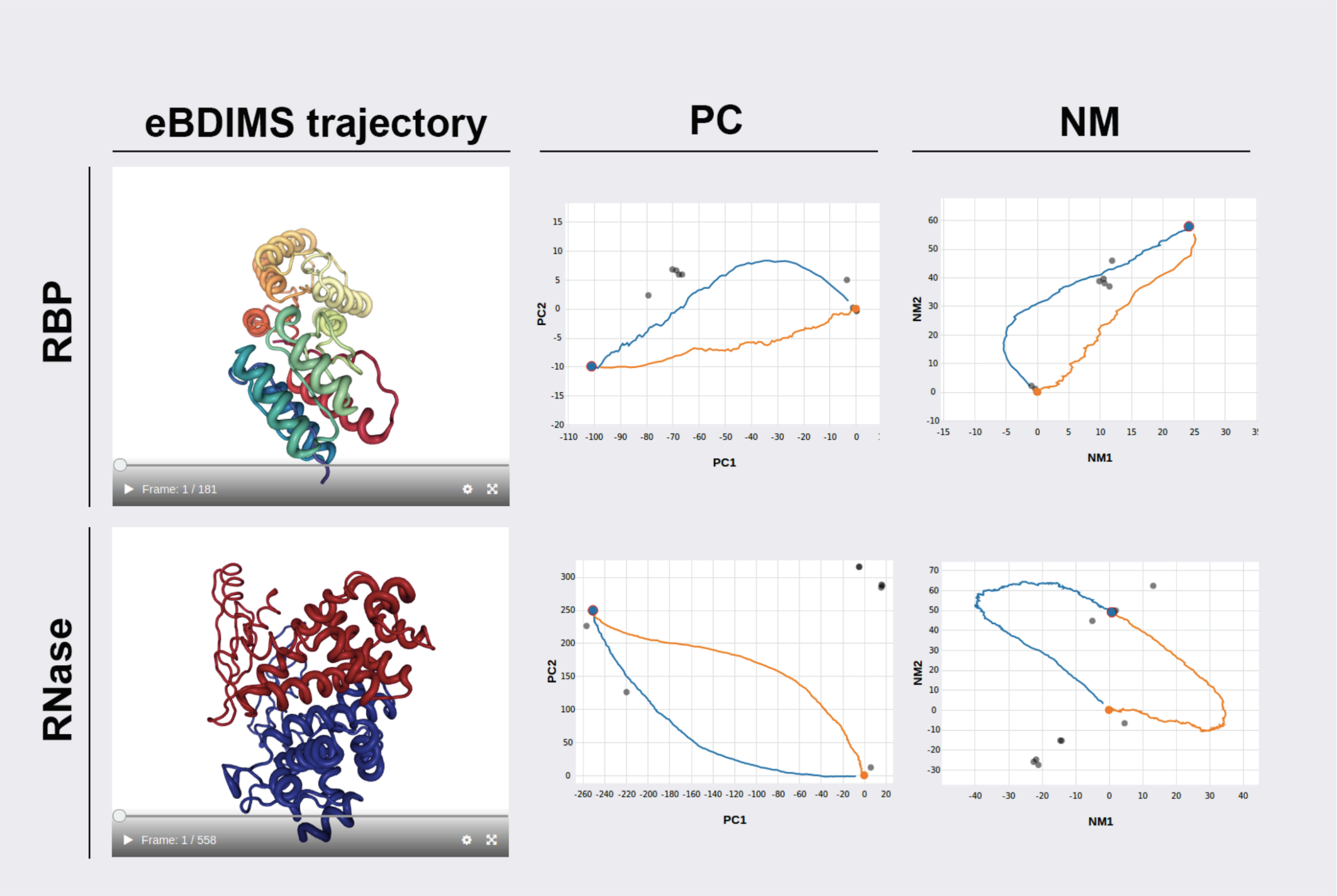

**Figure 2.** Screenshot of eBDIMS results for RBP (top) and RNaseIII (bottom) to illustrate the equivalence of the 2D-projection NM and PC partition for simple harmonic transitions like RBP hinge opening. NM-space and PC-space representations both project the transition intermediates as a cluster along the first motion axis, which is approached by the forward eBDIMS pathway (blue).

## SUPPLEMENTARY MATERIAL

**Table S1.** Summary of EBDIMS Benchmark Results available online

| Name | Start | Target | Nres (n-mer) | Initial rMSD | Final rMSD | EBDIMS Time | NMA* α (NM1), α (NM2) $\alpha_{10}$ | Nens | Variance PC1-PC2 |
|---|---|---|---|---|---|---|---|---|---|
| *RBP* | 1ba2 | 2dri | 270 (1) | 6.2 | 0.32<br>0.35 | 257s<br>163s | 0.83(1), 0.4(2)<br>0.97 | 11 | 97%-2% |
| *5NTase* | 1oid | 1hpu | 524 (1) | 9.3 | 0.35<br>0.34 | 797s<br>872s | 0,5(3), 0.3(8)<br>0.75 | 16 | 95%-4% |
| *RNaseIII* | 1yyo | 1yyw | 436(2) | 17.8 | 0.96<br>0.77 | 785s<br>652s | 0.4(9), 0.3(1)<br>0.70 | 11 | 51%-43% |
| *SERCA* | 2c9m | 1t5s | 994(1) | 14.16 | 0.89<br>0.75 | 6989s<br>7274s | 0.53(2), 0.52(3)<br>0.86 | 65 | 57%-28% |
| *GLIC* | 4npq | 4hfi | 1550(5) | 2.6 | 0.42<br>0.43 | 14400s<br>14400s | 0.73(4), 0.24(3)<br>0.79 | 46 | 42%-29% |
| *Calmodulin* | 1cll | 3ewv | 134(1) | 14.8 | 0.95<br>1.7 | 52s<br>69s | 0.51(4), 0.4(2)<br>0.85 | 44 | 42%-27% |

** Defined in Eq.5*